# Deep learning–based COVID-19 pneumonia classification using chest CT images: model generalizability

**Dan Nguyen**[1,2,a], **Fernando Kay**[3], **Jun Tan**[2], **Yulong Yan**[2], **Yee Seng Ng**[3], **Puneeth Iyengar**[2], **Ron Peshock**[3], **Steve Jiang**[1,2,a]

[1]Medical Artificial Intelligence and Automation (MAIA) Laboratory, University of Texas Southwestern Medical Center, Dallas, TX, USA
[2]Department of Radiation Oncology, University of Texas Southwestern Medical Center, Dallas, TX, USA
[3]Department of Radiology, University of Texas Southwestern Medical Center, Dallas, TX, USA
[a]Co-correspondence authors: {Dan.Nguyen, Steve.Jiang}@UTSouthwestern.edu

## Abstract

**Background:** Since the outbreak of the COVID-19 pandemic, research efforts around the world have begun to focus on the identification or categorization of COVID-19 positive patients using artificial intelligence (AI) technologies on various medical data. Some developed AI models have shown very high accuracy for diagnosis and prognosis of COVID-19 positive patients. However, concerns have been raised over the generalizability of these models, given the heterogeneous factors in training datasets such as patient demographics and pre-existing clinical conditions. The goal of this study is to examine the severity of this problem by evaluating the generalizability of deep learning (DL) classification models trained to identify COVID-19–positive patients on 3D computed tomography (CT) datasets from different countries.

**Methods:** We collected one internal dataset at UT Southwestern (UTSW) (337 patients), and three external datasets from three different countries: 1) CC-CCII Dataset (China), 2) COVID-CTset (Iran), and 3) MosMedData (Russia). We divided all the data into 2 classes: 1) COVID-19–positive and 2) COVID-19–negative patients. The total number of patients/scans we used for each data set was 101/101 (UTSW), 929/1544 (CC-CCII), 95/281 (COVID-CTset), and 856/856 (MOSMedData) for the positive labels, and 236/236 (UTSW), 1813/2634 (CC-CCII), 282/1068 (COVID-CTset), and 0/0 (MOSMedData) for the negative labels. We randomly divided the data, by patients, into 72% training, 8% validation, and 20% testing. We trained nine identical DL-based classification models by using various combinations of the datasets: 1) UTSW, 2) CC-CCII, 3) COVID-CTset, 4) UTSW + CC-CCII, 5) UTSW + COVID-CTset, 6) CC-CCII + COVID-CTset, 7) UTSW + CC-CCII + COVID-CTset, 8) CC-CCII + COVID-CTset + MosMedData, 9) UTSW + CC-CCII + COVID-CTset + MosMedData.

**Results:** The models trained on a single dataset achieved accuracy/area under the receiver operating characteristics curve (AUC) values of 0.87/0.826 (UTSW), 0.97/0.988 (CC-CCCI), and 0.86/0.873 (COVID-CTset) when evaluated on their own dataset. The models trained on multiple datasets and evaluated on a test set from one of the datasets used for training performed better. However, the performance dropped close to an AUC of 0.5 for all models when evaluated on a different dataset. For example, the model trained on the UTSW and COVID-CTset together achieved accuracies/AUCs to 0.91/0.937 and 0.94/0.926 when evaluated on the UTSW and COVID-CT-set datasets, respectively, but evaluating the model on the CC-CCII dataset yielded accuracy/AUC values of only 0.57/0.475. Including the MosMedData, which only contained positive labels, into the training did not necessarily help the performance on the other datasets.

**Conclusion:** The models could identify COVID-19–positive patients if the testing data were in the same dataset as the training data, and they performed better when trained on multiple datasets. However, we observed poor performance, close to random guessing, when evaluating the model on a dataset that it had never seen. Multiple factors likely contribute to these results, including but not limited to patient demographics, pre-existing clinical conditions, and differences in image acquisition or reconstruction, causing a data shift among different study cohorts.

# I. Introduction

Since the outbreak of the 2019 coronavirus disease (COVID-19) in December 2019, the total worldwide death count due to COVID-19 has exceeded a million deaths[1]. While COVID-19 can affect many organ systems and cause fever, flu-like symptoms, cardiovascular damage, and pulmonary injury. Since many of those who contract COVID-19 develop pneumonia, the most common presentation of COVID-19 is pneumonia. While some patients are asymptomatic or have mild symptoms, a small percentage of patients may develop severe acute respiratory distress syndrome (ARDS) that requires intubation in intensive care and is associated with poor prognosis. There is over 60% mortality once they progressed to a severe illness stage[2]. Since many chest CTs are performed for reasons other than pulmonary symptoms, an automated tool that can opportunistically screen chest CTs for the disease can potentially be used to identify patients with COVID-19. Firstly, it has been suggested that patients with COVID-19 when identified early stages can be treated to prevent progression to the later stage of the disease[3-5]. Secondly, identification of asymptomatic patients in early stages provide a time window when they can isolate themselves to prevent the spread to others.

Several efforts around the world have focused on the identification or categorization of COVID-19–positive patients according to their various types of medical data. As part of the effort to understand and control this disease, large COVID-19 datasets of different formats have been curated and publicly released around the world. One group of studies focuses on using artificial intelligence (AI) technologies, in particular deep learning (DL)–based models, to detect COVID-19 from chest radiography and computed tomography (CT). These studies found high accuracy rates ranging from 82% to 98%[6-16]. The high accuracy rates are promising and encourage the use of this technology in the clinical setting.

However, the generalizability of these models to other clinical settings around the world is not clear. The data usually found in clinical practice are often incomplete and noisy, and there may have high intra- and inter-study variability among different environments. This scenario often makes it difficulty from a research perspective to develop for algorithms and implement them in the clinic. Due to many restrictions on sharing patient data, many algorithms are developed with limited data that are specific to a clinic or a region. However, differences in several demographic factors, such as a populations distribution of race, ethnicity, and geography, can greatly impact the overall accuracy and performance of an algorithm in a different clinical setting[17]. In addition, different methods of data collection by hospitals around the world may also impact an algorithm's performance. Because the boom of AI technologies has only happened within the last several years, the number of studies testing the robustness and performance of AI algorithms across various clinical settings is extremely limited[17]. Therefore, there is very little knowledge about how well a model will perform in a realistic clinical environment over time.

For example, Barish et al.[18] demonstrated that a particular public model developed by Yan et al.[19] that predicted mortality from COVID-19–positive patients—which performed well on an internal dataset with an accuracy 0.878—failed to accurately predict mortality on an external dataset, with an accuracy of only around 0.5. Another similar negative study applied Yan et al.'s model on an external dataset and drew similar conclusions about the accuracy of its mortality prediction[20]. A systematic review of 107 studies with 145 prediction models was conducted, and they found that all models had high bias, due to non-representative control dataset and overly optimistic reported performance[21], which can additionally lead to unrealistic expectations among clinicians, policy makers, and patients[22]. An article by Bachtiger et al. had concluded that this boom of AI models for covid-19 focused far too much on developing novel prediction models without a comprehensive understanding of its practical application and biases from the dataset[23]. Others have similarly concluded that AI has yet to have any impact on the pandemic at hand, and that extensive and comprehensive gathering of diagnostic COVID-19–related data will be essential do develop useful AI models[24].

As part of the efforts to collect data, large datasets of 3D computed tomography (CT) scans with COVID-19–related labels have been publicly released. This provides us with an opportunity to study the generalizability of DL algorithms developed using these volumetric datasets. In this study, we collected and studied one internal dataset collected at UT Southwestern and three large external datasets from around the world: 1) China Consortium of Chest CT Image Investigation (CC-CCII) Dataset (China)[25], 2) COVID-CTset (Iran)[26], and 3) MosMedData (Russia)[27]. We trained a DL-based classification model on various combinations of the datasets and evaluated the model performance on held-out test data from each of the datasets.

## II. Methods

II.1. Data

We collected one internal dataset at UT Southwestern (UTSW) and three large datasets from around the world that are publicly available—1) China Consortium of Chest CT Image Investigation (CC-CCII) Dataset (China), 2) COVID-CTset (Iran), and 3) MosMedDat (Rusia)—which is summarized in Table 1. The UTSW dataset is composed of three subsets of anonymized imaging data obtained retrospectively. The study protocol was approved by the institutional review board and the requirement for informed consent was waived. The first subset includes patients who tested positive for Severe Acute Respiratory Syndrome Coronavirus-2 on real-time polymerase chain reaction between March and November 2020 and who had a chest CT scan performed within the first seven days of the diagnosis. All chest CT scans were obtained according to standard clinical care – common clinical indications were to assess worsening respiratory status and to rule out pulmonary thromboembolism. Chest CT is not obtained as a first line modality to diagnose or screen for COVID-19 at UTSW. The second and third subsets include patients who had a chest CT scan obtained as part of standard clinical care between March and May 2019, i.e., pre-COVID-19 pandemic phase. The radiologic reports of these studies were screened by a cardiothoracic radiologist with 12 year of clinical experience. The reports were labeled as having radiologic findings suggestive of infection or not. The adjudication was based on the presence of radiologic patterns usually associated with infection, including ground-glass opacities, consolidation, and nodular pattern, if such findings were described as fitting a differential diagnosis of infectious process based on the impression by the primary interpreting radiologists. These studies were consecutively selected to match the sex and age distribution of the COVID-19 positive subset and to represent two control groups with a balanced representation of chest CT showing findings suggestive of infection (118) and findings not related to infection

(118). The CC-CCII dataset was obtained from six different hospitals: 1) Sun Yat-sen Memorial Hospital and Third Affiliated Hospital of Sun Yat-sen University, 2) The first Affiliated Hospital of Anhui Medical University, 3) West China Hospital, 4) Nanjing Renmin Hospital, 5) Yichang Central People's Hospital, 6) Renmin Hospital of Wuhan University. The COVID-CTset dataset was from the Negin Medical Center, and the MOSMedData dataset was from municipal hospitals in Moscow, Russia.

For consistency in training and testing the models in our study, we divided all the data into two classes: 1) COVID-19–positive and 2) COVID-19–negative patients. Note that the MosMedData does not have conclusive negative–label patients, as CT-0 might contain both positive and negative patients. Accordingly, we omitted the CT-0 category from this study. Most scans in this study had a matrix size of 512 x 512 x n, where n was a variable number of slices. The small number of scans that had a reduced matrix size, images were linearly interpolated to match the 512 x 512 x n resolution.

Some data were available in Hounsfield Units (HU), while other data were available in relative intensity values (e.g., 0 to 255). Because the data formatting varied across datasets, we performed clipping and normalization operations. First, if the data were displayed in HU, we clipped the minimum number to be -1000 HU. For evaluation, the data were normalized from 0 to 1 prior to evaluation by the DL model. For training, multiple normalization methods were used as part of a data augmentation technique. The complete data augmentation is further described in section in Section II.3. Figure 1 shows example CTs of COVID-19 positive patients from each dataset.

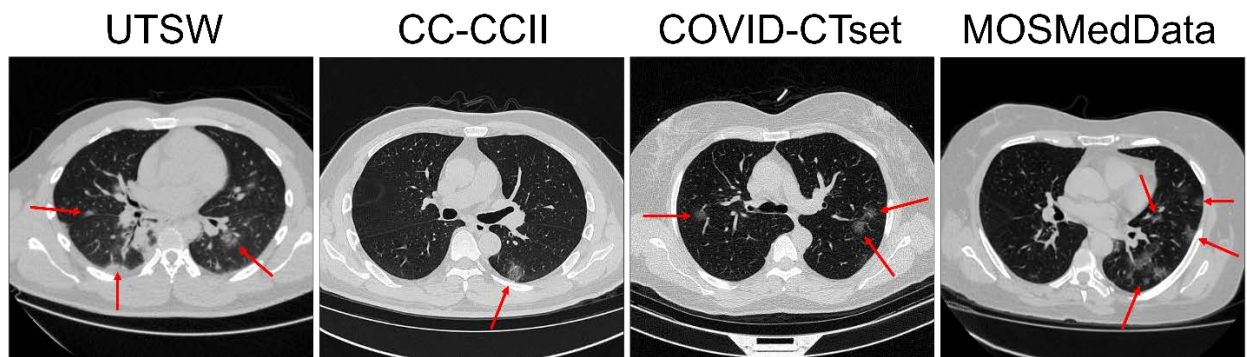

**Figure 1: Slice view of example CTs from each dataset. Red arrows show patchy ground-glass opacities with round morphology, which are typical findings in COVID-19 pneumonia.**

For training, validating, and testing the model, the positive labels of the UTSW dataset was randomly split into 73 train, 8 validation, and 20 test patients and scans (one 3D scan per patient). The positive labels of the CC-CCII dataset were randomly split into 669 train, 74 validation, and 186 test patients, or 1110 train, 122 validation, 312 test scans. The positive labels of the COVID-CTset were randomly split into 68 train, 8 validation, and 19 test patients, or 201 train, 23 validation, and 57 test patients. The positive labels of the MosMedData were randomly split into 616 train, 69 validation, and 171 test patients and scans (one 3D scan per patient. CT-0 category was omitted).

For the negative labels, the UTSW dataset was randomly split into 170 training, 18 validation, and 48 testing patients and scans (one 3D scan per patient). The CC-CCII dataset was randomly split 1305 train, 145 validation, and 363 test patients, or 1891 train, 203 validation, and 540 test scans.

The COVID-CTset was randomly split into 259 train, 29 validation, and 72 test patients, or 770 train, 84 validation, and 214 test scans.

**Table 1: Summary of data used in the study. These datasets include full volumetric CT scans of the patients**

| Dataset | Origin | Description | | | Available at: |
|---|---|---|---|---|---|
| | | # patients | # 3D scans | label | |
| UTSW | o UT Southwestern Medical Center | 101 | 101 | COVID-19 positive | *See footnote[1] |
| | | 118 | 118 | Infection (negative) | |
| | | 118 | 118 | Findings Unrelated to Infection (negative) | |
| China Consortium of Chest CT Image Investigation (CC-CCII) Dataset | o Sun Yat-sen Memorial Hospital and Third Affiliated Hospital of Sun Yat-sen University, Guangzhou, China <br> o The first Affiliated Hospital of Anhui Medical University, Anhui, China <br> o West China Hospital, Sichuan, China <br> o Nanjing Renmin Hospital, Nanjing, China <br> o Yichang Central People's Hospital, Hubei, China <br> o Renmin Hospital of Wuhan University, Wuhan, China | 929 | 1544 | COVID-19 positive | http://ncov-ai.big.ac.cn/download |
| | | 964 | 1556 | Common Pneumonia (negative) | |
| | | 849 | 1078 | Normal Lung (negative) | |
| COVID-CTset | o Negin Medical Center, Sari, Iran | 95 | 281 | COVID-19 positive | https://github.com/mr7495/COVID-CTset |
| | | 282 | 1068 | Normal lung (negative) | |
| MosMedData | o Municipal hospitals in Moscow, Russia | 254 | 254 | CT-0 – not consistent with pneumonia (can include both COVID-19 positive | https://mosmed.ai/ |

---

[1] UTSW dataset is non-public. In accordance with HIPAA policy, access to the dataset will be granted on a case by case basis upon submission of a request to the corresponding authors and the institution.

| | | 684 | 684 | and negative) |
| --- | --- | --- | --- | --- |
| | | | | CT-1 – Mild (COVID-19 positive) |
| | | 125 | 125 | CT-2 – Moderate (COVID-19 positive) |
| | | 45 | 45 | CT-3 – Severe (COVID-19 positive) |
| | | 2 | 2 | CT-4 – Critical (COVID-19 positive) |

## II.2. Model Architecture

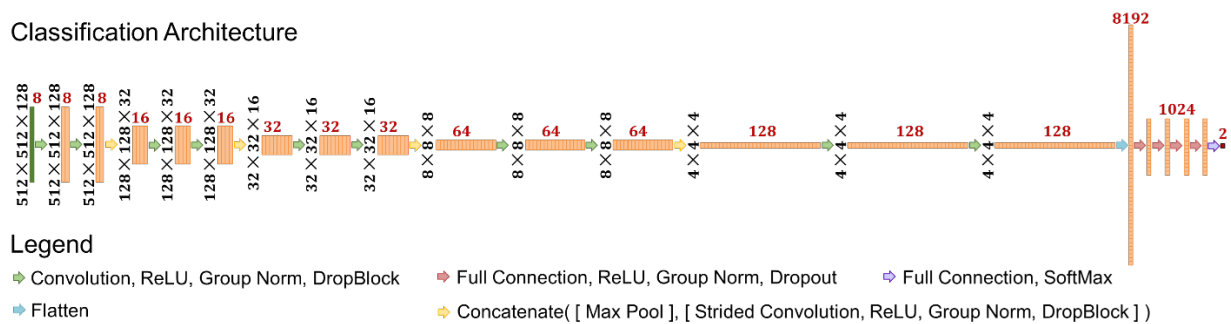

**Figure 2: Schematic of deep learning architecture used in the study. Black numbers represent the feature shape of each layer prior to the flattening operation. Red numbers represent the number of features at each layer.**

The model used in this study was a classification style convolutional neural network (CNN) model[28-31], with specifics shown in Figure 2. The input shape was set to 512 x 512 x 128. In total, there are 5 resolution levels of convolutions and 4 downsampling operations prior to the flattening operation. The downsampling size also varied each time, and was set as (4,4,4), (4,4,2), (4,4,2), and (2,2,2), respectively. In total, this converts the data shape from 512 x 512 x 128 to 4 x 4 x 4. At each resolution level, a series of operations consisting of Convolution, Rectified Linear Unit activation (ReLU), Group Normalization[32], and DropBlock[33] is applied twice, consecutively. The convolution kernel size varied at each resolution level: (3,3,3), (5,5,5), (5,5,3), (5,5,3), and (3,3,3), respectively. The number of filters, indicated by the red numbers in Figure 2, at each convolution started at eight, and doubled after each downsampling operation. After these operations, the feature data are flattened into a single vector of length 8192. Then, a series of operations consisting of fully connected calculations, ReLU, Group Normalization, and Dropout[34] follows. This is performed at total of four times, calculating 1024 features each time. Then, one more full connection is applied to reduce the data into two outputs, and a softmax operation is applied.

III.3. Training and Data Augmentation

In total, nine models were trained in this study using the training and validation data outlined in Section III.1. and were split into two categories: 1) single dataset training and 2) multiple dataset training. We trained three models on a single dataset, one each on the UTSW, CC-CCII, and the COVID-CTset datasets. No model was trained on the MosMedData by itself, since this dataset does not have any negative labels. For multiple-dataset training, we trained 6 models with different combinations of datasets: 1) UTSW + CC-CCII, 2) UTSW + COVID-CTset, 3) CC-CCII + COVID-CTset, 4) UTSW + CC-CCII + COVID-CTset, 5) CC-CCII + COVID-CTset + MosMedData, and 6) UTSW + CC-CCII + COVID-CTset + MosMedData.

Some additional operations were applied to format and augment the CT data for model training. For CT data with less than 128 slices, slices of zeros were padded onto the CT slices until the total data volume had 128 slices. The number of slices superior and inferior to the CT data was uniformly and randomly decided at each iteration. For data with more than 128 slices, a random continuous volume of 128 slices was selected. The data were then normalized in one of two ways: 1) from 0 to 1, or 2) from 0 to $\frac{max(data)}{2^n}$, where $n$ is the smallest integer possible while keeping $2^n$ larger than the maximum value in the CT volume. The normalization method was randomly chosen with a 50% chance during each training iteration. An additional step was applied to decide, at a 50% chance, whether this data would be fed into the model for training, or if additional data augmentation would be applied. If yes to additional data augmentation, then the function randomly flipped, transposed, rotated, or scaled the data. For the flip augmentation, there was a 50% chance it would individually apply a flip to each axis (row, column, slice). For the transpose augmentation, there was a 50% chance it would transpose the row and column of the data (no transpose operation was ever applied using the slice dimension). For the rotate augmentation, a random integer, {0,1,2,3}, was generated and multiplied against 90° to determine the rotation angle, then applied only on the row and column dimensions. For the scale augmentation, there was a 50% chance that a scaling factor was applied, and the scale was a uniform random number from 0 to 1.

Each model trained for a total of 250000 iterations, using the Adam optimizer[35] with a learning rate of $1 \times 10^{-5}$. To prevent overfitting on the training data, the accuracy was evaluated on the validation data every 500 iterations, and the instance of the model with the highest validation accuracy was saved as the final model for evaluation. The models were trained using NVIDIA V100 GPUs with 24 GB of memory.

III.4. Evaluation

All nine of the trained models were evaluated on the test data of each dataset. For volumes with less than 128 slices, zero padding on the slices was evenly applied in the superior and inferior directions, to keep the data centered. For volumes greater than 128 slices, a sliding window technique was applied across the volume and the model made multiple predictions. The number of slices in a patch was 128 and the stride size was 32 slices. The prediction with the highest COVID-19 probability was taken as the model's final prediction.

A threshold was selected based on maximizing the prediction accuracy on the validation data and applied to the testing set. In the cases where the "optimal" threshold was a trivial value (e.g., threshold = 0 for the MosMedData, which only has positive labels), we took the argmax of the output as the prediction instead. The true positives ($TP$), true negatives ($TN$), false positives ($FP$), and false negatives ($FN$) were counted, and a normalized confusion matrix is generated for each dataset. Averaged confusion matrices were calculated with and without the MosMedData. An evenly weighted average was chosen.

Receiver operating characteristic (ROC) curves were calculated on the test data by varying the positive predictive threshold from 0 to 1, at 0.01 intervals. The area under the curve (AUC) was calculated to determine the overall performance of each model on each dataset. The MosMedData was excluded from the ROC and AUC analysis, since it is missing negative labels.

## III. Results

Each model took about five days on average to train on a GPU. In total, for nine models, this is equivalent to 45 GPU-days of training. Each model prediction takes an average of 0.53 seconds, which makes it very useful for near real-time application.

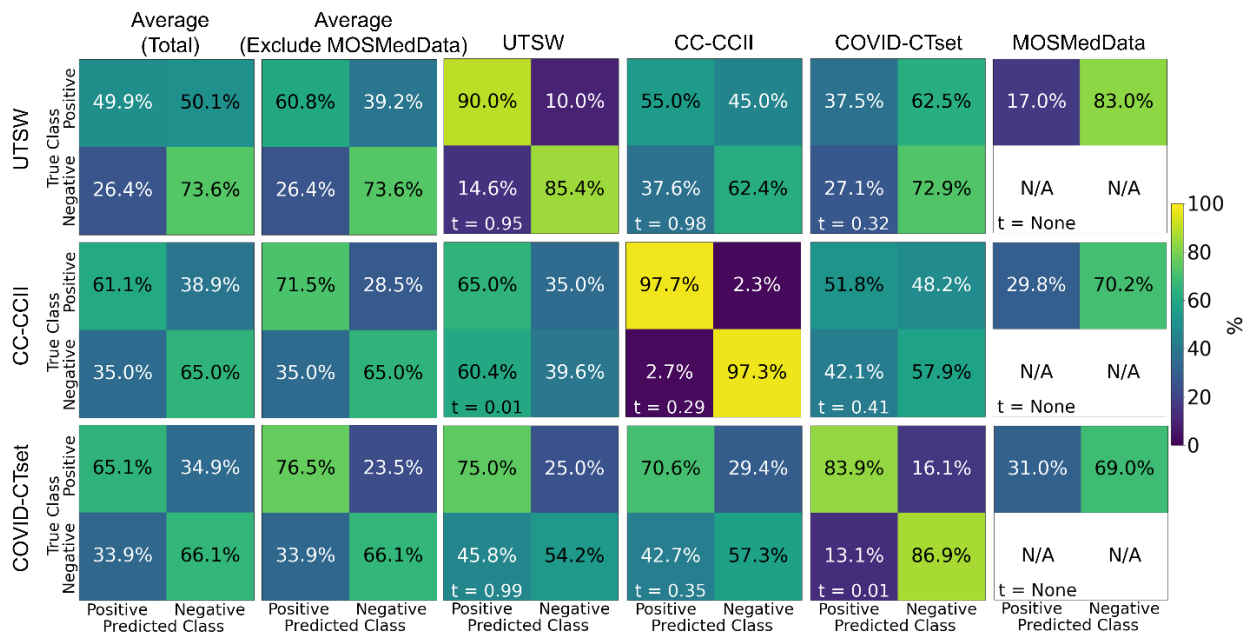

**Figure 3: Confusion matrices on the test data for each of the models trained on a single dataset. Each row represents the datasets that the model was trained on, and each column represents the datasets that the model was evaluated on. Note that MosMedData does not have any negative label data. The labeling threshold used for each model is indicated on the lower left of each confusion matrix ($t = \#$); this threshold was fine-tuned from the validation data to maximize the trace of the confusion matrix for each respective category. $t = None$ indicates that the argmax of the model output was used to determine whether the prediction was positive or negative.**

The single dataset models' predictive accuracy $\left(\frac{TP+TN}{TP+TN+FP+FN}\right)$ on the test dataset is displayed in Figure 3. Overall, each model performed best on the dataset that it trained on, with an accuracy as high as 0.97 for the CC-CCII model evaluated on the CC-CCII data. The model that performed the worst on its own dataset was COVID-CTset, with an accuracy of 0.86. The UTSW model had an accuracy of 0.87 on its own dataset. Since the test data were held out of the training and validation phase, it is a strong indicator that the model did not overfit to its specific training data. However, the models performed much more poorly when evaluated on a dataset they had not seen before, which signifies that the model did not generalize well to the new dataset type. The worst performance was the CC-CCII model evaluated on the UTSW dataset, which had an accuracy of 0.47. All three models had poor performance on the MosMedData dataset and classified the majority of the patient cases as negative.

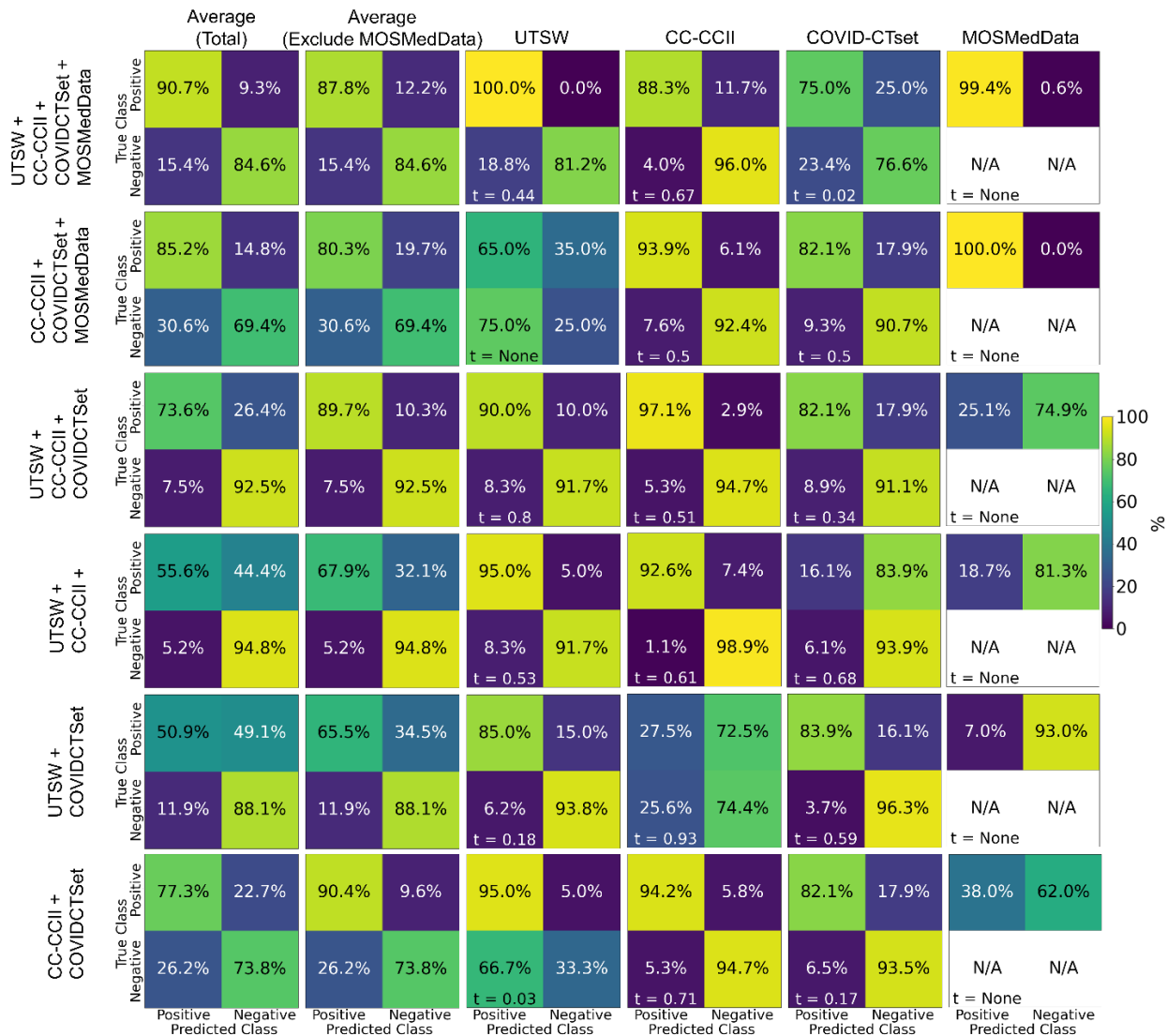

**Figure 4: Confusion matrices on the test data for each of the models trained on multiple datasets. Each row represents the datasets that the model was trained on, and each column represents the datasets that the model was evaluated on. The labeling threshold used for each model is indicated on the lower left of each confusion matrix ($t = \#$). $t = None$ indicates that the argmax of the model output was used to determine whether the prediction was positive or negative.**

Figure 4 shows the confusion matrices of the performance of models trained on multiple datasets against the test data. The multiple dataset model that had the best accuracy when evaluated on the UTSW test set was the UTSW + CC-CCII model, with 0.93 accuracy. When evaluating on the CC-CCII test set, there was a tie for best accuracy 0.96 between the UTSW + CC-CCII + COVID-CTset and the UTSW + CC-CCII models. When evaluating on the COVID-CTset, the UTSW + COVID-CTset performed best, with a 0.94 accuracy. The best multiple-dataset models outperformed their single-dataset counterparts with regards to accuracy. However, these models still had poor accuracy when evaluated on a test dataset they have not seen before. For example, the model trained with the UTSW and COVID-CTset together had improved accuracies to 0.91 and 0.94 when evaluated on the test sets of the UTSW and COVID-CTset datasets, respectively. However, when evaluated on the CC-CCII dataset, the accuracy was 0.57. Including the MosMedData in the model training improved the total average performance but did not improve the performance when evaluating models on the individual UTSW, CC-CCII, and COVID-CTset datasets.

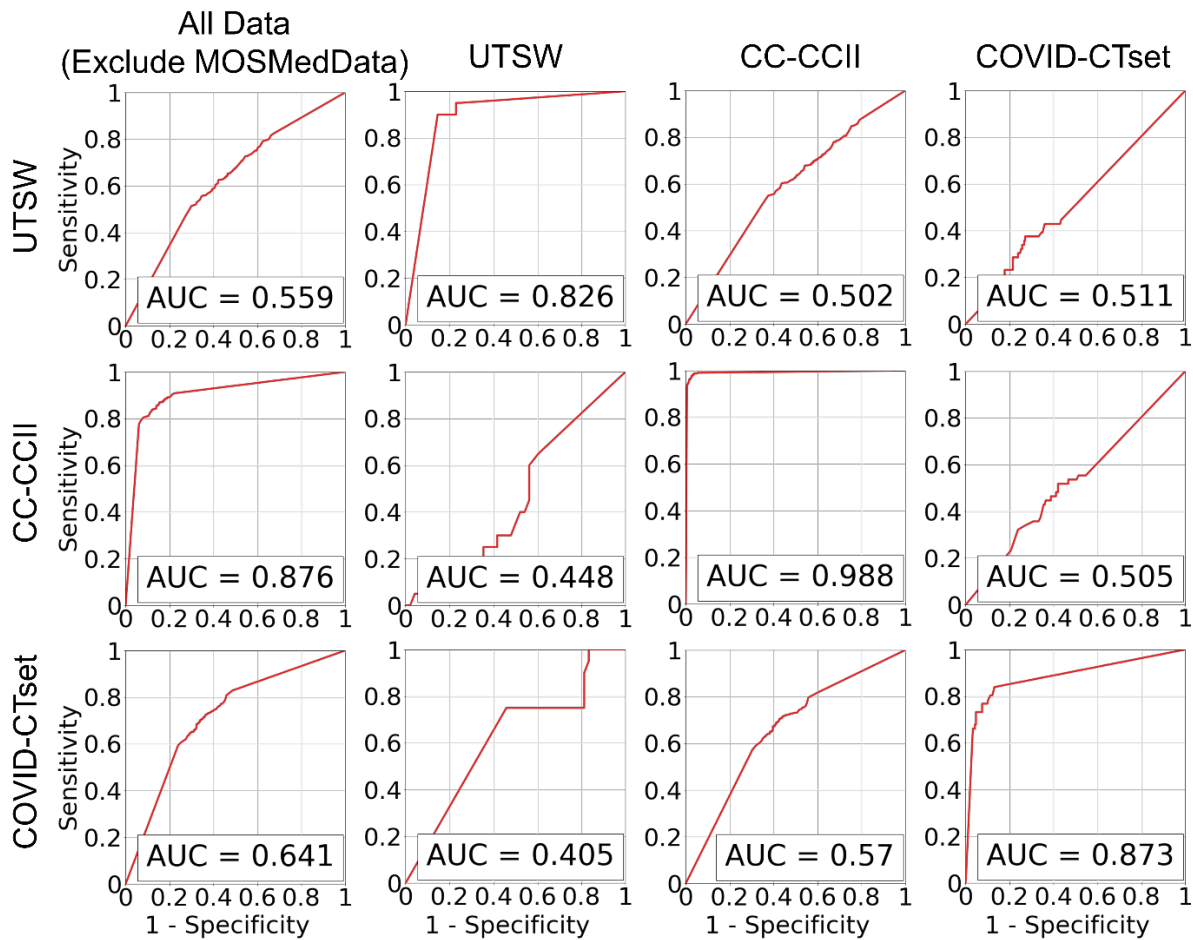

**Figure 5: ROC curves on the test data for the models trained on single datasets. Each row represents the datasets that the model was trained on, and each column represents the datasets that the model was evaluated on.**

Figure 5 shows the ROC curves of the single-dataset models. The models, when evaluated on the same dataset that they trained on, showed good AUCs of 0.826 (UTSW), 0.988 (CC-CCII), and 0.873 (COVID-CTset). The models performed considerably worse when evaluated on different datasets, with AUCs ranging from 0.405 to 0.570, which is close to just random guessing (i.e., AUC=0.5). The ROC curves of the multiple-dataset models are shown in Figure 6. For each dataset—UTSW, CC-CCII, and COVID-CTset—the best performing models were the UTSW + COVID-CTset (AUC = 0.937), the UTSW + CC-CCII + COVID-CTset (AUC = 0.989), and the UTSW + COVID-CTset (AUC = 0.926) models, respectively. Since the test data were held entirely separate from the model development process, and used only for evaluation, this shows once again that the models did not overfit to their own training data. Similar to the single-dataset models, the multiple-dataset models also performed poorly when predicting on datasets they had never seen before, with AUCs ranging from 0.380 to 0.540.

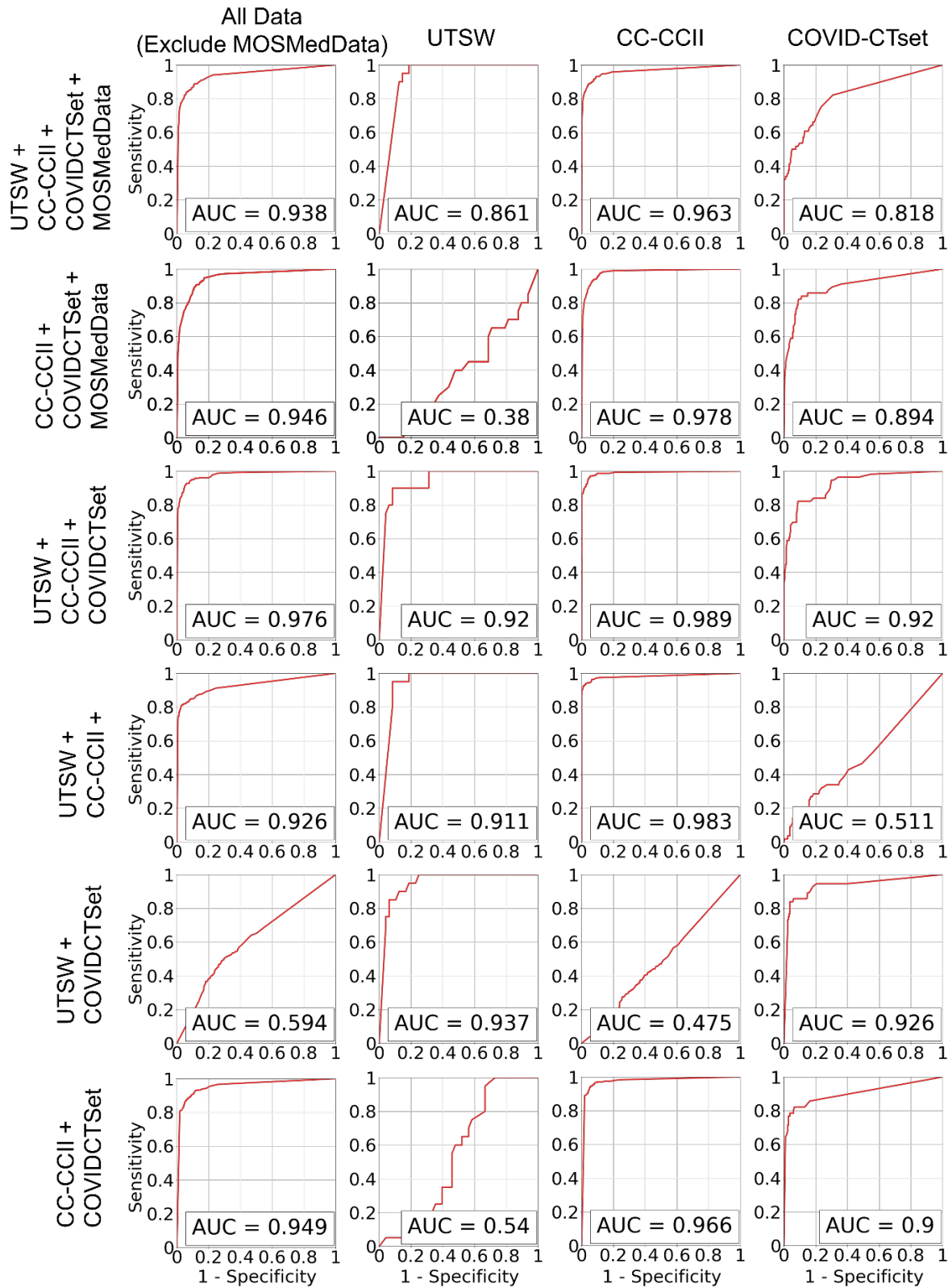

**Figure 6: ROC Curves on the test data for the models that trained on multiple datasets. Each row represents the datasets that the model was trained on, and each column represents the datasets that the model was evaluated on.**

## IV. Discussion

In this study we demonstrate that our DL models can correctly identify patients that are COVID-19–positive with high accuracy, but only when the model was trained on the same datasets that it was tested on. Otherwise, the performance is poor—close to random guessing—which indicates that the model cannot easily generalize to an entirely new dataset distribution that it has never seen before for COVID-19 classification. Several data augmentation techniques were applied during training to prevent overfitting on the test set. In addition, the weights of the model that performed the best on the validation data with regards to accuracy were used as the final model. Dropout and DropBlock regularization were added to further prevent the model from overfitting.

We additionally observed that certain combined dataset models performed best for particular datasets in detecting patients who are positive for COVID-19. For example, we found that the highest performing model in the dataset from UTSW dataset was obtained when the training step combined UTSW and COVID-CTset datasets. This may have occurred due to the relatively low sample count in the UTSW dataset (73 positive, 170 negative patients for training); therefore, adding data samples from COVID-CTset improved with DL-model's AUC from 0.826 to 0.937 on the UTSW dataset. However, adding more data from different distributions into the training did not always monotonically improve the model's performance. For example, adding the CC-CCII data for training did not improve the model performance, with the AUC of 0.920 for the UTSW dataset. Adding MosMedData into the training lowered the performance of the model on the other three datasets. This is likely because the original intent of MosMedData was to train a model to categorize the severity of COVID-19 into five classes and, therefore lacked negative labels. Without definitive negative labels, our models likely learned simply to identify the data source as MosMedData and compromised some of their learning capacity and performance to use the relevant imaging features for the predictions. This does serve as an important lesson in data collection: datasets from a particular healthcare center or region should be fully representative of the task at hand to be used in training. Simply collecting COVID-19–positive patients from one source and negative patients from a different source is likely to introduce an uncorrectable bias during training that led to poor model performance.

Although this study did not fully explore possible techniques to improve robustness and prevent overfitting, it may serve as a baseline for future model generalization studies that use medical data for the clinical implementation of COVID-19–related classification models. We will continue to explore the limits of model generalization with respect to improving the algorithm and to the intra- and inter-source data variability, regarding the identification of COVID-19–positive patients by their medical data. As a whole, the deep learning models achieved high performance on the unseen test set from the same distribution that they were trained on, which indicates that we did not have typical overfitting problem with the training data. The low performance on datasets that the models had never seen before may actually be an indicator that the problem is not in the approach to initial algorithm development—the problem may be the transfer and deployment of the algorithm to a new clinical setting. Creating a globally generalizable algorithm is a tall order when people around the world have vastly different demographics and data collection protocols. With limited data and learning time, these AI algorithms are bound to fail when they encounter a unique data distribution they have never seen before. This results underscore the limited versatility of AI algorithms which may hamper widespread adoption of AI algorithms for automated diagnosis of radiology images. This is in contrast to radiologists who in general can easily adapt to new clinical practices quickly. Perhaps we need to recalibrate our mindset in regards to the expectation for these AI algorithms—we should expect that these AI algorithms will always need to be fine-tuned to the local distribution when implemented and deployed in a specific clinical setting, then need to be retuned over time as distributions inevitably shift, either through demographic shifts or the advancement of new treatment technologies. Transfer learning and continuous learning techniques[36] are an active field of research, and may become critical components to rapidly transferring, deploying, and maintaining an AI model into the clinic.

AI tools designed for automatic identification of diseases on CT datasets, such as COVID-19, will only succeed if they can prove their robustness against a wide array of patient populations, scan protocols, and image quality. Notwithstanding, they hold the promise of becoming a powerful resource for identifying diseases where time to detection is a critical variable. In the case of COVID-19, it is well known that many infections are asymptomatic, of which up to 54% will present abnormalities on chest CT[37]. Thus, COVID-19 can be incidentally found on routine imaging. Timely identification of incidental cases of COVID-19 on chest CT by AI tools could lead to adequate prioritization of scans for reporting, resulting in prompt initiation of disease tracking and control measures. Moreover, the model architecture developed in this work can also serve as a template for similar tools tailored for detecting other clinical conditions.

## V. Conclusion

The deep learning models were capable of identifying COVID-19–positive patients when the testing data was in the same dataset as the training data, whether the model was trained on a single dataset or on multiple datasets. However, we found poor performance, close to random guessing, when models were evaluated on datasets that they had never seen. This is likely due to different factors, such as patient demographics, image acquisition methods/protocols, or diagnostic methods, causing a data shift between different countries' data. This lack of generalization for the identification of COVID-19–positive patients may not necessarily mean that the models were trained poorly, but rather the distribution of the training data may be too different from the evaluation data. Transfer learning and continuous learning may become imperative tools for tuning and deploying a model in a new clinical setting.

## Acknowledgements

We would like to thank Jonathan Feinberg for editing the manuscript.

## Dataset and Code Availability

UTSW dataset is non-public. In accordance with HIPAA policy, access to the dataset will be granted on a case-by-case basis upon submission of a request to the corresponding authors and the institution.

The DL models and related code developed in this study are available upon request for non-commercial research purposes.